\documentclass[12pt,preprint]{aastex}

\def\ifundefined#1{\expandafter\ifx\csname#1\endcsname\relax}

\def\la{\mathrel{\hbox{\rlap{\hbox{\lower4pt\hbox{$\sim$}}}\hbox{$<$}}}}
\def\ga{\mathrel{\hbox{\rlap{\hbox{\lower4pt\hbox{$\sim$}}}\hbox{$>$}}}}

\newcommand{\be}{\begin{eqnarray}}
\newcommand{\ee}{\end{eqnarray}}

\ifundefined{ensuremath}\def\ensuremath#1{\relax\ifmmode{#1}}
\else${#1}$\fi\else\relax\fi
\ifundefined{nuc}\def\nuc#1#2{\relax\ifmmode{}^{#1}{\protect\textrm{#2}}
\else${}^{#1}$#2\fi}\else\relax\fi

\newcommand{\kmps}{\ensuremath{\mathrm{km}~\mathrm{s}^{-1}}}

\newcommand{\msol}{\ensuremath{{\mathrm{M}_\odot}}}

\newcommand{\nni}{\ensuremath{^{56}\mathrm{Ni}}}

\def\ang{\ensuremath{\mathrm{\AA}}}

\newcommand{\phx}{\texttt{PHOENIX}}

\shortauthors{Baron, E. et~al.}
\shorttitle{Spectral Modeling of a 3-D Pulsating Reverse
  Detonation}

\begin{document}

\title{Detailed Spectral Modeling of a 3-D Pulsating Reverse
  Detonation Model: Too Much Nickel}

\author{ 
  E.~Baron,\altaffilmark{1,2}
 David J.~Jeffery,\altaffilmark{1,3}  David
Branch,\altaffilmark{1} Eduardo
Bravo,\altaffilmark{4,5} Domingo Garc\'ia-Senz,\altaffilmark{4,5}  and  Peter 
H.~Hauschildt\altaffilmark{6}
 }

 \altaffiltext{1}{Homer L.~Dodge Department of Physics and Astronomy,
   University of Oklahoma, 440 West Brooks, Rm.~100, Norman, OK
   73019-2061; USA, baron,branch,jeffery@nhn.ou.edu}

\altaffiltext{2}{Computational Research Division, Lawrence Berkeley
  National Laboratory, MS 50F-1650, 1 Cyclotron Rd, Berkeley, CA
  94720 USA}

\altaffiltext{3}{Department of Physics, University of Idaho,  PO Box
  440903,  Moscow, ID 83844-0903}

\altaffiltext{4}{Departament F\'isica i Enginyeria Nuclear,
  Universitat Polit\`ecnica de Catalunya, Diagonal 647, 08028
  Barcelona, Spain; eduardo.bravo,domingo.garcia@upc.edu}

\altaffiltext{5}{Institut d'Estudis Espacials de Catalunya, Barcelona, Spain}

\altaffiltext{6}{Hamburger Sternwarte, Gojenbergsweg 112,
21029 Hamburg, Germany; yeti@hs.uni-hamburg.de}

\begin{abstract}
  We calculate detailed NLTE synthetic spectra of a Pulsating Reverse
  Detonation (PRD) model, a novel explosion mechanism for Type Ia
  supernovae. While the hydro models are calculated in 3-D, the
  spectra use an angle averaged hydro model and thus some of the 3-D
  details are lost, but the overall average should be a good
  representation of the average observed spectra. We study the model
  at 3 epochs: maximum light, seven days prior to maximum light, and 5
  days after maximum light. At maximum the defining Si~II feature is
  prominent, but there is also a prominent C~II feature, not usually
  observed in normal SNe~Ia near maximum. We compare to the early
  spectrum of SN 2006D which did show a prominent C~II feature, but
  the fit to the observations is not compelling. Finally we compare to
  the post-maximum UV+optical spectrum of SN~1992A. With the broad
  spectral coverage it is clear that the iron-peak elements on the
  outside of the model push too much flux to the red and thus the
  particular PRD realizations studied would be intrinsically far
  redder than observed SNe~Ia. We briefly discuss variations that
  could improve future PRD models.
\end{abstract}


\section{Introduction} 

Understanding the nature of the explosion mechanism of Type Ia
supernovae (SNe~Ia) is important for the use of SNe Ia as precision
cosmological probes as well as for stellar evolution and
nucleosynthesis.   Determining something about the nature of the
explosion will almost certainly shed light on the physical explanation
of the empirical light curve shape--luminosity relationship
\citep{philm15,rpk95,rpk96,philetal99,goldhetal01,KW07} as well as further
physical parameters associated with the observed diversity in the
SN~Ia population. Understanding the physical nature of the diversity is
crucial in being able to estimate systematic errors that may arise in
comparing the characteristics of the nearby sample with the
cosmological one. 

While there is widespread agreement that the SN~Ia progenitor
consists of a C+O white dwarf that accretes material from a companion
and ignites the C+O fuel when the Chandrasekhar mass is reached, the
exact mechanism of explosion is not known.  The explosion could take the
form of a super-sonic detonation (shock-wave), or a sub-sonic
deflagration. If a unexpanded Chandrasekhar mass WD detonates, the
resulting products would be almost entirely $\nni$ and
$\alpha$-particles \citep{ATW71}, contrary to the signature intermediate mass
elements of silicon, sulfur, and calcium that define the  SNe~Ia
class. For many years the standard SN~Ia model has been the 1-D,
parameterized, deflagration model W7 \citep{nomw7}. In that model the
speed of the flame was adjusted in order to reproduce the
approximately correct amount of $\nni$, as well as intermediate mass
elements. Nevertheless, the central regions of the model undergo
significant electron capture and far too much neutron-rich material is
produced, polluting the interstellar medium. \citet{brach00} showed that with pf-shell model
calculations the electron capture rates are reduced over those of
standard sd-shell rates of \citeauthor{FFN80} (\citeyear{FFN80,FFN82a,FFN82b,FFN85}),
we are unaware of hydrodynamical calculations that show that the
problem is completely solved.  In order to avoid some of the
problems of W7, the delayed-detonation model has gained significant
favor \citep{gko05,gko04a,K93c,khok91a,khok91c}. In this scenario, the
ignition begins as a deflagration, allowing the star to pre-expand,
before the deflagration transforms into a detonation at lower
densities, which produces both $\nni$ and intermediate mass
elements. While this process is known to occur in confined terrestrial
situations, it is not known whether it can take place in the
unconfined white dwarf.

The last decade has seen the beginning of full 3-D hydro calculations
\citep{RHN02a,RHN02b,RNH02,hrn00,RHN99,RHNKG99,Gamezoetal03,gko04a,gko05,GSB05,BGS03a,GSB03a,GSetal99,rhnw06,roepkeetal06,RH05a,RH05b,roepke05}.
Much of the work has focused on pure deflagration models as well as
developing the necessary sub-grid techniques to handle the essentially
infinite dynamic range (due to turbulence) or studying the effects of
differing numbers of ignition points. The initial ignition conditions
may well be unknowable since about one hundred years before ignition
the white dwarf becomes turbulent \citep{rwh07}. Recently, there has
been general agreement that pure deflagration models leave behind 
significant amounts of unburnt carbon and oxygen and that the kinetic
energy of the explosion is too low to reproduce the observed spectra
of normal SNe~Ia \citep{gko04a,hoef_right07}. The alternative is a
delayed detonation
\citep{K93c,khok91a,khok91c,khok91b,gko05,hoef_right07} where the
deflagration accelerates into a detonation or some other process  that turns
the deflagration into a detonation.
Two alternatives that act
 in a similar manner to the delayed-detonation scenario are the
 ``Gravitationally Confined Detonation'' \citep{PCL04,jordangcd07} and
 the ``Pulsating Reverse Detonation'' \citep{bravo06}. 

 We present detailed NLTE synthetic spectral calculations of an angle
 averaged version of PRD5.5 a pulsating-reverse-detonation
 calculation (E.~Bravo \& D.~Garc\'ia-Senz, in preparation). The
 detailed chemical composition was determined from the hydro model
 with a post-processing code that included a maximum of 725 isotopes.
 The evolution of the electron fraction $Y_e$ was followed along with
 the hydrodynamic calculation with the aid of tables of electron
 capture rates in NSE matter. The procedure was identical to that used
 in \citet{GSetal99}.

\section{Calculations}

The calculations were performed using the multi-purpose stellar
atmospheres program \phx~{version \tt 15}
\citep{hbjcam99,bhpar298,hbapara97,phhnovetal97,phhnovfe296}. Version
15 incorporates many changes over previous versions used for supernova
modeling \citep{bbh07,bbbh06} including many more species in the
equation of state (83 versus 40), twice as many atomic lines, and many
more species treated in full NLTE.
\phx\ solves the radiative transfer equation along characteristic rays
in spherical symmetry including all special relativistic effects.  The
non-LTE (NLTE) rate equations for many ionization states are solved
including the effects of ionization due to non-thermal electrons from
the $\gamma$-rays produced by the  radiative decay of $^{56}$Ni, which
is produced in the
supernova explosion.  The atoms and ions calculated in NLTE are: H~I,
He~I--II, C~I-III,  O~I-III, Ne~I, Na~I-II, Mg~I-III, Si~I--III,
S~I--III, Ca~II, Ti~II, Fe~I--III, Ni~I-III, and Co~I-III. These are all the
elements whose features make important contributions to the observed
spectral features in SNe~Ia.

Each model atom includes primary NLTE transitions, which are used to
calculate the level populations and opacity, and weaker secondary LTE
transitions which are included in the opacity and implicitly
affect the rate equations via their effect on the solution to the
transport equation \citep{hbjcam99}.  In addition to the NLTE
transitions, all other LTE line opacities for atomic species not
treated in NLTE are treated with the equivalent two-level atom source
function, using a thermalization parameter, $\alpha =0.10$
\citep{snefe296}.  The 
atmospheres are iterated to energy balance in the co-moving frame;
while we neglect the explicit effects of time dependence in the
radiation transport equation, we do implicitly include these effects,
via explicitly including the rate of gamma-ray deposition in the generalized
equation of radiative equilibrium and in the rate equations for the
NLTE populations.

The outer boundary condition is the total bolometric luminosity in the
observer's frame. The inner boundary condition is that the flux at the
innermost zone ($v=1500$~\kmps) is given by the diffusion equation.
We treat the $\gamma$-ray deposition by solving the full radiative
transfer for a grey opacity.
Detailed fitting of the observed spectra determines all the parameters.

\phx\ has been well tested on SNe~Ia
\citep{nug1a95,nugseq95,nughydro97,l94d01,bbbh06} and particularly on
SN~1994D \citep{l94d01,bbbh06}. \phx\ is subjected to rigorous
regression testing with each new version.

\section{The Model}

In the Pulsating Reverse Detonation model \citep{bravo06} the
deflagration starts in a number of rising plumes that are already
somewhat close to the surface. These plumes burn without much
expansion of the white dwarf and hence burn to iron peak elements. The
energy released by this initial deflagration is not enough to unbind
the star and the outer parts fall back producing an accretion
shock. The unburnt core settles back into equilibrium and the
accretion shock detonates the outer parts which are at low enough
density that they burn to intermediate mass elements.  Thus the
composition structure in this model is not completely stratified and
the outer layers are polluted by iron and radioactive nickel. The
precise amount of polluting Fe-group elements is sensitive to the
outcome of the deflagrative phase of the explosion,  more efficient
subsonic burning reduces the amount of stable Fe-group pollution. The mass
fraction of 
radioactive material close to the surface of the ejecta depends as
well on the amount of electron captures that, in turn, are a sensitive
function of the central density at ignition. 

We present results from model PRD5.5 (E.~Bravo \& D.~Garc\'ia-Senz,
in preparation). The model started from five sparks, incinerating 0.14
\msol\ during the initial subsonic combustion phase, and later on a
detonation processed most of the remaining carbon and oxygen. Finally,
the explosion produced $1.3 \times 10^{51}$~ergs of kinetic energy,
and ejected 0.86~\msol\ of \nni, 0.15 \msol\ of intermediate-mass
elements, and 0.18 \msol\ of carbon and oxygen.  In model PRD5.5 the
ejected mass of stable iron and nickel was $0.1 \msol$, most of which was
moving at velocities in excess of 10,000~\kmps. The final averaged
distribution of elements in velocity space is quite similar to that of
model PRD6 \citep[see Fig.~3 in][]{bravo06}. 

\begin{figure}
\centering
\includegraphics[width=0.85\textwidth]{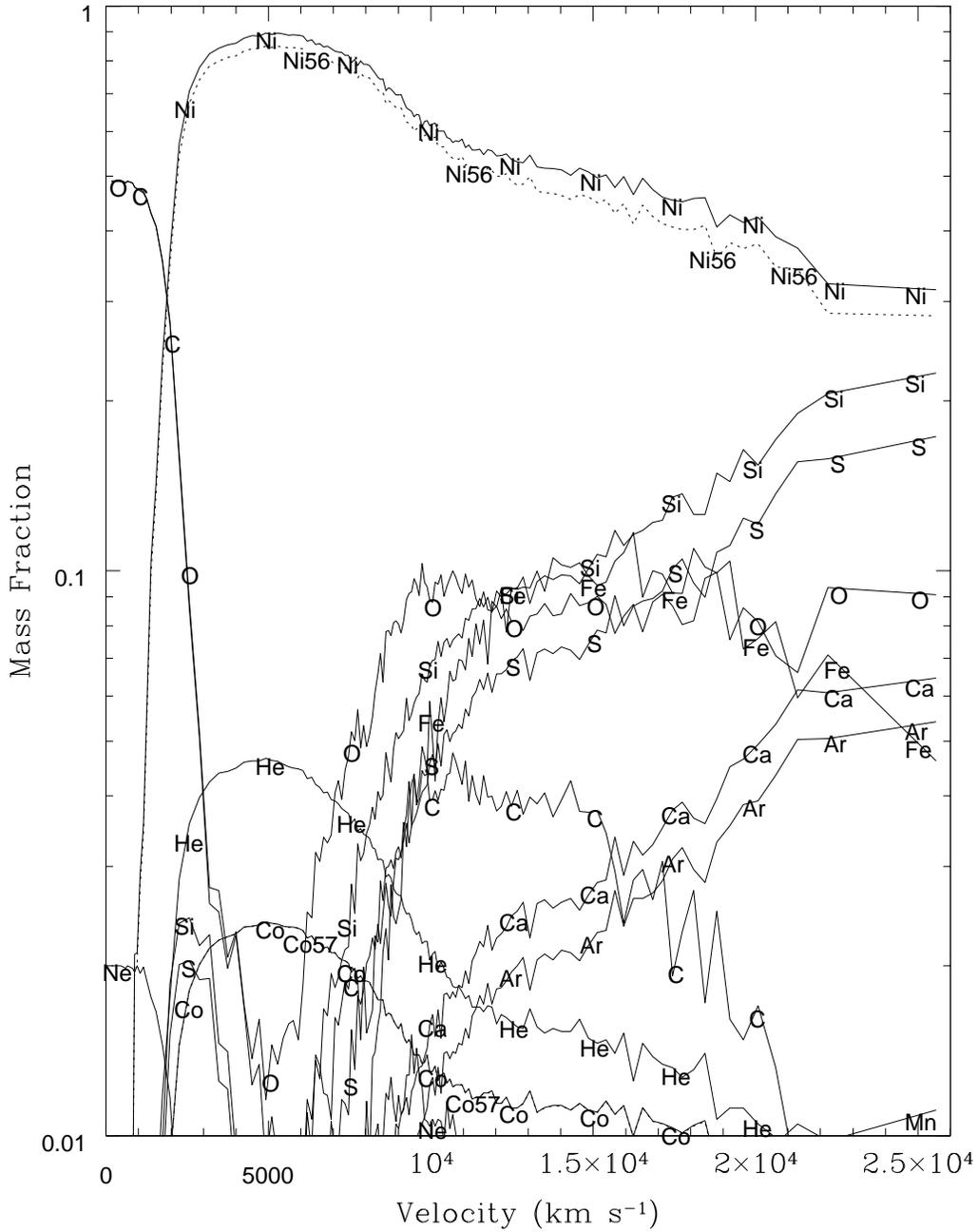}
\caption{The angle averaged composition of the PRD5.5 at
time zero after the explosion: i.e., before significant radioactive
decay has occur. Elements are denoted by solid curves; radioactive
nuclides by dotted curves. \label{fig:comps}}
\end{figure}

The 3-D 
composition of model PRD5.5 is quite clumpy.  The angle averaged
version is rather 
uniform in the outer layers above $\sim 8000$~\kmps, where
the supernova spectrum mostly forms before a few days past maximum
light (Fig.~\ref{fig:comps}).  While the outer layers are dominated by
$^{56}$Ni,  
there is significant silicon, sulfur, and oxygen as required for
SN Ia spectra.  It is a quantitative question that only a code such
as \phx\ can determine whether or not the outer $^{56}$Ni conflicts
with observations.  There is also significant carbon in the
outer layers and C+O dominates the innermost region.  Synthetic
spectrum modeling by \citet{kozma05} of the deflagration model
{c3\_3d\_256\_10s} \citep{roepke05} found that abundant C+O matter in
the interior was inconsistent with late-time nebular spectra.  Model
{c3\_3d\_256\_10s} had more interior C+O than PRD5.5, therefore
future nebular modeling of PRD5.5 is also important.

\section{Results}

\begin{figure}
\centering
\includegraphics[width=0.70\textwidth,angle=90]{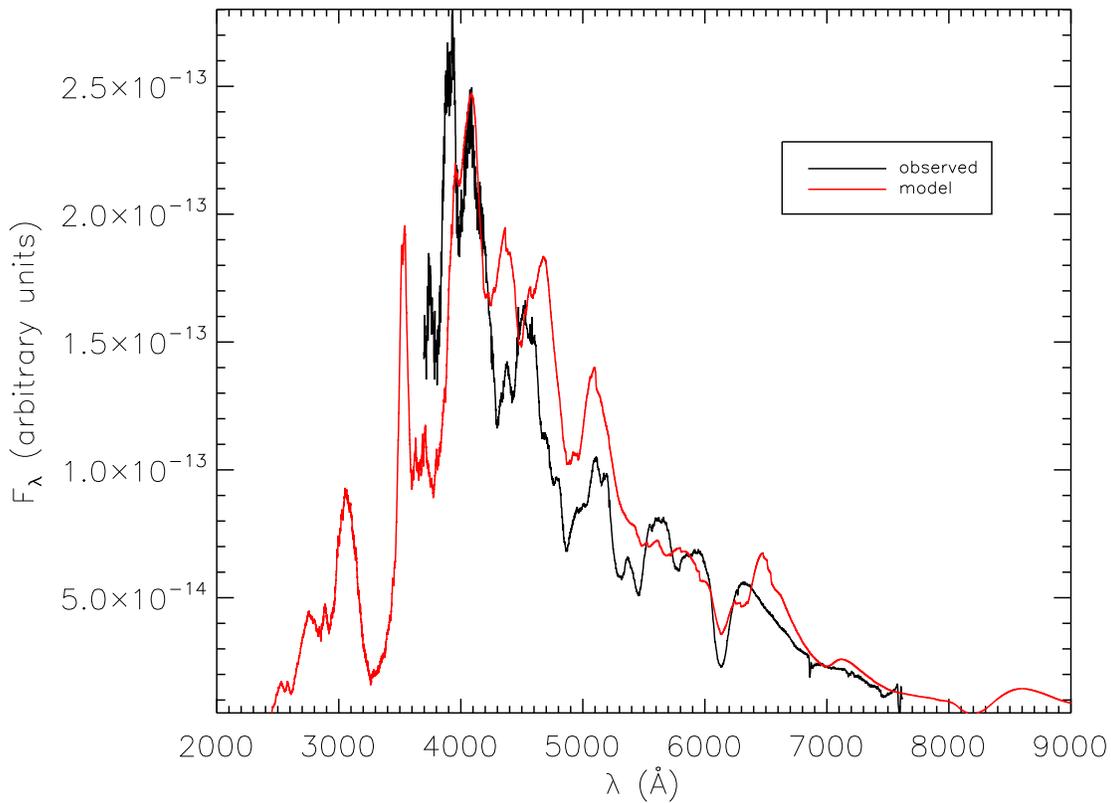}
\caption{
  \label{fig:sn94d_max} The synthetic spectra of a full NLTE model of
  PRD5.5, 20 days after explosion, is compared to the
  observed spectrum of SN~1994D on March 21, 1994 (the time of B
  maximum). The observed
  spectrum has been corrected for redshift assuming a velocity of
  448~\kmps\ and a reddening of $E(B-V) = 0.06$. 
}
\end{figure}

We first examine the maximum light spectrum of SN~1994D, which we have
used in previous papers to study the detailed spectral behavior since
it is nearby and extremely well-observed
\citep{branchcomp105,l94d01}. Figure~\ref{fig:sn94d_max} shows the
results of our best-fit model (we use the outer boundary condition of
the total bolometric luminosity in the observer's frame to find the
best fit). The overall fit is not all that bad, the feature at
6150~\ang\ attributed to Si~II $\lambda 6355$ is reasonably well fit,
however the ``W'' feature attributed to S~II is totally washed
out. Additionally just to the red of the 6150 feature is another
feature almost certainly due to C~II $\lambda 6580$. This feature has
in fact been observed in early SNe~Ia spectra
\citep{branch98aq03,thomas06d07}. While the 6150~\ang\ feature is the
usual defining feature for SN~Ia, the S~II ``W'' is even a more
reliable diagnostic. SNe~Ib/c spectra do show features that have been
identified as Si~II, but only SNe~Ia show the S~II feature. Since
silicon and sulfur are made nearly simultaneously one might expect the
lines to appear together, however the Si~II line is much stronger than
the S~II lines, thus it requires more sulfur to obtain the S~II ``W''
feature.  Empirically S~II seems to be the best classifier of
SNe~Ia. Even the extremely odd SN~Ia 2002ic which shows strong narrow
Balmer lines in emission shows solid evidence for the S~II ``W''
\citep{ham02ic03}.  If S~II is used as the defining feature of SN~Ia,
then the synthetic spectrum of Fig.~\ref{fig:sn94d_max} is not that of
a SN~Ia. Since the angle averaging will automatically lead to some
washing out of the flux spectrum we should be careful not to draw too
strong conclusions about the absence of S~II lines, but their complete
absence in the presence of Si~II is interesting. While the relative
amount of silicon and sulfur in the model is quite similar S~III is
less abundant than Si~III only by a factor of two or more everywhere
in the outer atmosphere, thus it is unclear why there is no evidence
for S~II in the synthetic spectra (the lines of the ionization stage
just below the most abundant one tend to be strongest, since there is
then energy to excite the levels above the ground state). We speculate
that the complete absence of evidence for S~II in the synthetic
spectrum is a peculiarity of the particular NLTE state realized in the
outer ejecta of PRD5.5.  If so (and only a very detailed decomposition
of the spectrum formation in future work can tell), only a \phx\ 
level calculation could reveal this peculiarity inherent in PRD5.5.

\begin{figure}
\centering
\includegraphics[width=0.70\textwidth,angle=90]{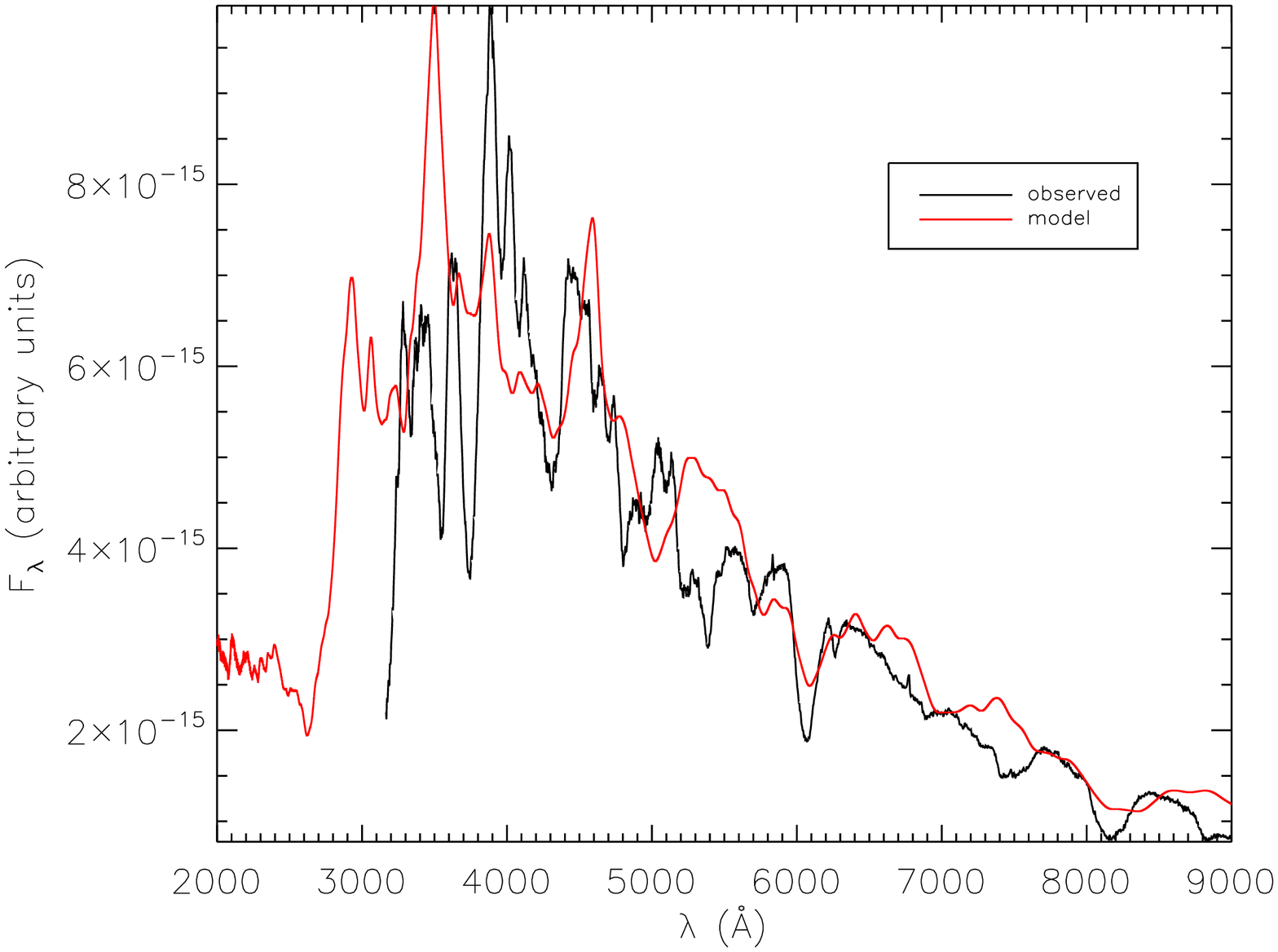}
\caption{
  \label{fig:sn06D_m07} The synthetic spectra of a full NLTE model of
  PRD5.5, 13 days after explosion, is compared to the
  observed spectrum of SN~2006D on 7 days prior to B
  maximum. The observed
  spectrum has been corrected for redshift $z=0.00853$. No reddening
  correction was made.
}
\end{figure}

Figure~\ref{fig:sn06D_m07} shows the best fit spectrum compared to
SN~2006D seven days prior to $B$ maximum, where the C~II $\lambda6580$
line is prominent in the observed spectrum. At this earlier time the
emission portion of the carbon line in the synthetic spectrum is
evident but the absorption 
trough is not blue enough and does not have the right shape. The
triangular shape in the observed spectrum is almost certainly due to
line blending. S~II is again missing in the synthetic spectrum. 
The outermost
compositions used in the spectral modelling are somewhat uncertain due to
lower resolution in the outer parts of the spectral grid and to the need to
interpolate from the gridless hydro models
(the grid needed for
good resolution in synthetic spectra is different from that needed in
hydro). This effect, coupled with the 
smearing created by the angle averaging procedure would lead us not to expect
perfect agreement with the observed lineshapes. 

\begin{figure}
\centering
\includegraphics[width=0.70\textwidth,angle=90]{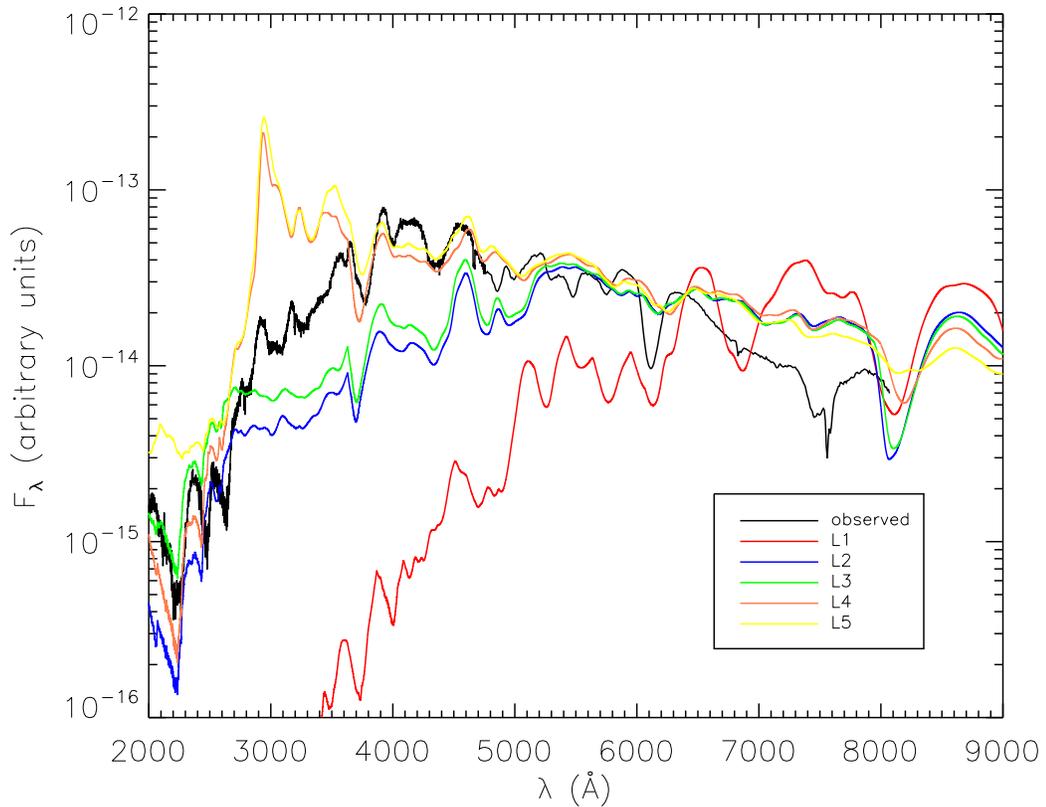}
\caption{
  \label{fig:sn92A_d25} The synthetic spectra of a full NLTE model of
  PRD5.5, 25 days after explosion, are compared to the
  observed spectrum of SN~1992A observed five days past B
  maximum. The indicated lines show synthetic spectra of models with
  increasing total 
  bolometric luminosity in the observer's frame (L1-L5). The observed
  spectrum has been corrected for redshift assuming a velocity of
  1845~\kmps. No reddening
  correction was made. While the optical spectra can be somewhat fit,
  no variation in the luminosity can fit the entire wavelength range
  from the UV to the optical.
}
\end{figure}

One of the key features of the pulsating-reverse-detonation model is
that it creates a composition inversion. That is, the initial
deflagration burns at high density and thus tends to make iron-peak
elements on the outside at high velocity. High velocity features have
been detected in supernova spectra at early times, but they tend to be
either unburnt material such as carbon, or intermediate mass elements
like calcium. \citet{branchcomp206} used high velocity Fe~II to fit
near maximum light supernova spectra, using the simple parameterized
code SYNOW. However, the optical depths used were low, and thus the
material could be primordial. It is impossible to use SYNOW to
directly determine abundances.  Figure~\ref{fig:sn92A_d25} shows the
combined \emph{HST} 
and optical spectrum of SN~1992A five days past maximum light
\citep{kir92a}. Plotted on this figure are five different choices for
the outer boundary condition, the total luminosity in the observer's
frame. The luminosity varies from the
lowest value to the highest by a 
factor of 13. The logarithmic bolometric luminosity (ergs~s$^{-1}$)
is: L1=-17.02,  
L2=-18.58,
L3=-19.07,
L4=-19.31,
L5=-19.81.
Clearly no value chosen for
the luminosity can reproduce 
the observed colors across  the UV+optical spectrum. This is
almost certainly due to the large amount of iron-peak elements in the
outer parts of the model. These elements have such strong line
blanketing that they always push much of the flux to the red and thus
the red colors are too red for all reasonable luminosities.

\section{Conclusions}

The pulsating-reverse-detonation model is a novel mechanism that
combines the advantages of the delayed detonation mechanism (higher
explosion energy, low amounts of unburnt fuel remaining) without
requiring a deflagration to detonation transition in an unconfined
medium. However, the mechanism still has to be explored further and
different variations will change the total mass in the initial
deflagration. We have examined just one particular realization here
and we have also studied only the angle averaged flux spectra
calculated in spherical symmetry, thus losing some of the important
3-D effects of the model. We note that we have also studied the
spectra of PRD6.0.  PRD6.0 burned carbon more efficiently during the
deflagration phase, thus the mass fraction of stable iron and nickel
in the outer layers of that model is nearly one order of magnitude
lower than in PRD5.5. However, the mass fraction of \nni\ in the
outermost layer is only slightly lower ($\sim 15$\%) in PRD6.0.  The
overall differences in the synthetic spectra were not
significant. This suggests that the composition inversion inherent in
the model will lead to supernova spectra and light curves that are
extremely red, which has not been observed. A way out of this would be
if there were some pre-expansion during the initial deflagration phase
so that the burning is to intermediate mass elements rather than all
the way to the iron-peak.  The pre-expansion is dependent on details
of the initial model. For instance, if a shallow enough thermal
gradient were present in the bubbles, the nuclear burning would
propagate initially with a shockless supersonic phase velocity that
would lead to pre-expansion before the combustion settled in a
subsonic flame regime. Another possibility is that the explosion takes
place in the core of a rotating white dwarf.  Since for the same
central density, fast rotators are more prone to synthesize more
intermediate-mass elements and less iron than static initial
models. The determination of the ignition configuration of the white
dwarf is one of the key problems in the physics of SNIa during the
explosion phase. Thus, we reiterate that the problem found in the
models studied here do not rule out the PRD scenario, just the
particular realizations that we have studied.  Clearly there is more
work to be done in studying this particular explosion mechanism. Our work
also shows the power of quantitative spectroscopy in validating
hydrodynamical explosion models.

\acknowledgements 
We thank the referee for a constructive report that improved the
presentation. 
This work was supported in part by NASA grant
NNG04GB36G, US DOE Grant DE-FG02-07ER41517, and NSF grants AST-0506028 and AST-0707704.  E.~Bravo and
D.G.-S. have been 
supported by the MEC grants AYA2005-08013-C03, AYA2004-06290-C02, by
the European Union FEDER funds and by the Generalitat de Catalunya.
E.~Baron acknowledges support from the
US Department of Energy Scientific Discovery through Advanced
Computing program under contract DE-FG02-06ER06-04.
This research used resources of the National Energy Research
Scientific Computing Center (NERSC), which is supported by the Office
of Science of the U.S.  Department of Energy under Contract
No. DE-AC03-76SF00098; and the H\"ochstleistungs Rechenzentrum Nord
(HLRN).  We thank all these institutions for a generous allocation of
computer time.



\end{document}